\documentclass[12pt]{article}

\usepackage{graphicx}
\usepackage{amsmath}
\usepackage{amssymb}

\def\CR{\nonumber \\}
\def\eq#1{(\ref{#1})}
\def\[#1\]{\begin{align}#1\end{align}}
\def\mH{{\mathcal H}}
\def\mJ{{\mathcal J}}
\def\mD{{\mathcal D}}
\def\mN{{\mathcal N}}
\def\mO{{\mathcal O}}
\def\pder#1{\frac{\partial}{\partial #1}}
\begin{document}

\begin{titlepage}
\title{
\hfill\parbox{4cm}{ \normalsize YITP-15-3 \\ WITS-MITP-001}\\
\vspace{1cm} 
Renormalization procedure for random tensor networks
and the canonical tensor model}
\author{
Naoki {\sc Sasakura}$^a$\thanks{\tt sasakura@yukawa.kyoto-u.ac.jp}
\and Yuki {\sc Sato}$^b$\thanks{\tt Yuki.Sato@wits.ac.za}
\\[15pt]
$^a${\it Yukawa Institute for Theoretical Physics, Kyoto University,}\\
{\it Kyoto 606-8502, Japan} \\
\\
$^b${\it National Institute for Theoretical Physics, }\\
{\it School of Physics and Mandelstam Institute for Theoretical Physics,} \\
{\it University of the Witwartersrand, Wits 2050, South Africa}
}
%\date{\today}
%\normalsize}
\maketitle
\thispagestyle{empty}
\begin{abstract}
\normalsize
We discuss a renormalization procedure for random tensor networks,
and show that the corresponding renormalization-group flow is given by the Hamiltonian vector flow
of the canonical tensor model, which is a discretized model of quantum gravity.
The result is the generalization of the previous one
concerning the relation between the Ising model on random networks and 
the canonical tensor model with $N=2$.
We also prove a general theorem which relates
discontinuity of the renormalization-group flow and the phase transitions
of random tensor networks.
\end{abstract}
\end{titlepage}

%%%%%%%%%%%%%%%%%%%%%
\section{Introduction}
\label{sec:intro}
%%%%%%%%%%%%%%%%%%%%%
Wilson's renormalization group \cite{Wilson:1974mb, Wilson:1973jj} 
is an essential and pedagogical tool in modern theoretical physics. 
Once a renormalization-group flow in a parameter space is given, 
one can read off relevant degrees of freedom at each step of coarse graining 
through change of parameters, and understand the phase structure in principle. 
Therefore, 
a renormalization-group flow gives us a quantitative and qualitative 
picture of a system concerned. 
The aim of this paper is to define a renormalization procedure 
and derive the corresponding flow equation for random tensor networks, 
in particular for those proposed as Feynman-graph expressions 
\cite{Sasakura:2014zwa,Sasakura:2014yoa}, 
through the use of the canonical tensor model (CTM, for short).      

First of all, 
CTM has been introduced by one of the authors as a model of quantum gravity 
by considering space-time as a dynamical fuzzy 
space \cite{Sasakura:2011sq,Sasakura:2012fb,Sasakura:2013gxg}.
CTM is a tensor model in the canonical formalism, which has
a canonical conjugate pair of rank-three tensors, 
$M_{abc}$, $P_{abc}$ ($a,b,c=1,2, \cdots,N$), as dynamical variables. 
This interpretation of tensorial variables in terms of a fuzzy space is different from the one made 
by original tensor models.
Historically, tensor models have been introduced as models of simplicial quantum gravity
in dimensions higher than two 
\cite{Ambjorn:1990ge,Sasakura:1990fs,Godfrey:1990dt}; 
although the original tensor models have some drawbacks, 
tensor models as simplicial quantum gravity 
are currently in progress as colored tensor 
models \cite{Gurau:2009tw,Gurau:2011xp} (See \cite{Gurau:2013pca}-\cite{Delepouve:2014hfa} for 
recent developments.).
In CTM, $N$, the cardinality of the rank-three tensors, may be interpreted as the number 
of ``points" forming a space, 
while physical properties of space-time such as dimensions, locality, etc. must emerge from 
the collective dynamics of these ``points."
So far, the physics of the small-$N$ CTM is relatively well understood: 
the classical dynamics of the $N=1$ CTM agrees with the minisuperspace 
approximation of general relativity in arbitrary dimensions \cite{Sasakura:2014gia}; 
the exact physical states have been obtained for $N=2$ in the full 
theory \cite{Sasakura:2013wza,Narain:2014cya} 
and for $N=3$ in an $S_3$-symmetric subsector \cite{Narain:2014cya}; 
intriguingly, physical-state wavefunctions, at least for $N=2,3$, 
have singularities where symmetries are enhanced \cite{Narain:2014cya}. 
However, similar brute-force analysis as above for $N > 3$ seems technically difficult 
because of the huge number of degrees of freedom of the tensorial variables, 
although in order to capture, for instance, emergence of space-time from CTM, 
the large-$N$ dynamics is supposed to be important. 
Thus, for the purpose of handling large-$N$ behaviors of CTM, 
the present authors have proposed the conjecture that statistical systems on random 
networks \cite{Sasakura:2014zwa,Sasakura:2014yoa,revnetwork}, 
or \textit{random tensor networks}, are intimately related to CTM \cite{Sasakura:2014zwa}:
the phase structure of random tensor networks is equivalent to what is implied by 
considering the Hamiltonian vector flow of CTM as the renormalization-group flow of random tensor networks. 
This conjecture has been checked qualitatively for $N=2$ \cite{Sasakura:2014zwa}. 
In fact, as more or less desired, random tensor networks turn out to be useful to find 
physical states of CTM with arbitrary $N$:
some series of exact physical states for arbitrary $N$ have been found as integral 
expressions based on random tensor networks \cite{Narain:2014cya}. 

In this paper, we prove the fundamental aspect of the above conjecture:
we show that the Hamiltonian vector flow of CTM can be regarded as 
a renormalization-group flow of random tensor networks for general $N$. 
Here the key ingredient is that the Lagrange multipliers of the Hamiltonian vector flow are determined by 
the dynamics of random tensor networks in the manner given in this paper.  
This is in contrast with the previous treatment for $N=2$, in which the Lagrange multipliers are 
given by a ``reasonable" choice \cite{Sasakura:2014zwa}.
In fact, the previous treatment turns out to have some problems for general $N$, as being discussed in this paper.

This paper is organized as follows. 
In section \ref{sec:previous}, we review CTM and random tensor networks.
We argue our previous proposal \cite{Sasakura:2014zwa} 
on the relation between CTM and random tensor networks, and its potential problems. 
In Section~\ref{sec:renorm}, we propose a renormalization procedure for random tensor networks 
based on CTM, and derive the corresponding renormalization-group flow.
In Section~\ref{sec:comp}, we compare our new and previous proposals with 
the actual phase structures of random tensor networks for $N=2,3$. 
We find that the new proposal is consistent with the phase structures, while the previous one is not.
In Section~\ref{sec:fixed}, we discuss the asymptotic behavior of the renormalization-group flow,
and clarify the physical meaning of the renormalization parameter.
In Section~\ref{sec:theorem}, we provide a general theorem 
which relates discontinuity of the renormalization-group flow and the phase transitions
of the random tensor networks.
Section~\ref{sec:discussion} is devoted to summary and discussions.     

%%%%%%%%%%%%%%%%%%%%%%%%%%%%%%%%%%%%%%%
\section{Previous proposal and its problems}
\label{sec:previous}
%%%%%%%%%%%%%%%%%%%%%%%%%%%%%%%%%%%%%%%
In this paper, we consider a statistical system \cite{ Sasakura:2014zwa,Sasakura:2014yoa}
parameterized by a real symmetric three-index tensor
$P_{abc}\ (a,b,c=1,2,\ldots, N)$.\footnote{In this paper, the tensor variable of the statistical system 
is denoted by $P$ for later convenience,
instead of $M$ used in the previous papers \cite{ Sasakura:2014zwa,Sasakura:2014yoa}.}
Its partition function is defined 
by\footnote{For later convenience, the normalizations of the partition function and $\phi$ are taken differently from those in the previous papers \cite{Sasakura:2014zwa,Sasakura:2014yoa}. 
This does not change the physical properties of the statistical system.}
\[
Z_n(P)=\int d\phi \ (P \phi^3)^n \, e^{-n \phi^2},
\label{eq:znm}
\]
where we have used the following short-hand notations,
\[
\int d\phi\equiv  \prod_{a=1}^N \int_{-\infty}^\infty d\phi_a, \ \ \
P\phi^3\equiv P_{abc}\phi_a \phi_b \phi_c, \ \ \
\phi^2\equiv \phi_a \phi_a.
\]

\begin{figure}
\begin{center}
\includegraphics{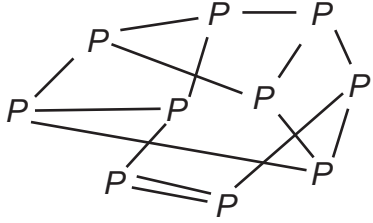}
\caption{A tensor network of $n=10$ trivalent vertices.}
\label{fig:network}
\end{center}
\end{figure}

By using the Wick theorem, the Gaussian integration of $\phi$ in \eq{eq:znm} can be evaluated by  
the summation over the pairwise contractions of all the $\phi$'s in $(P\phi^3)^n$.
Then the partition function \eq{eq:znm} 
can graphically be represented by the summation over all the possible 
closed networks of $n$ trivalent
vertices. In each of such networks, every vertex is weighted by $P_{abc}$, 
and the indices are contracted according to the connection of the network, 
as in Fig.\ref{fig:network}. 
Therefore, since the pairwise contractions are taken over all the possible ways, 
the statistical system represented by \eq{eq:znm} can be regarded as random tensor networks 
of $n$ trivalent vertices. In general, such a network may contain disconnected sub-networks,
but this probability vanishes in the thermodynamic limit 
$n\rightarrow\infty$.\footnote{
This graphical property can be checked by considering a sort of ``grand" 
partition function $Z(P)$ 
given by a formal sum of \eq{eq:znm} over $n$ with $n$-dependent weights, 
and comparing explicitly the perturbative 
expansions in $P$ of $Z(P)$ and those of $\log Z(P)$ for $N=1$. The former corresponds to 
the sums of the networks which may contain disconnected sub-networks, 
while the latter to connected networks only. See \cite{Sasakura:2014zwa} for details.}

For example in $N=2$, \eq{eq:znm} gives the partition function
of the Ising model on random networks \cite{ Sasakura:2014zwa,Sasakura:2014yoa}, if one takes
\[
P_{abc}=\sum_{i=1}^2 R_{ai}R_{bi}R_{ci} e^{H \sigma_i},
\]
where $\sigma_i$ represents the spin degrees of freedom taking $\sigma_{1}=1,\, \sigma_2=-1$, 
$H$ is a magnetic field, and $R$ is a two-by-two matrix satisfying 
\[
(R^T R)_{ij}=e^{J \sigma_i \sigma_j},
\label{eq:rr}
\]
with $J$ giving the nearest neighbor coupling of the Ising model. 
For a ferromagnetic case $J>0$, there exists a real matrix $R$ satisfying \eq{eq:rr}.

The partition function \eq{eq:znm} is obviously invariant under the orthogonal 
transformation $L\in O(N)$, which acts on $P$ as
\[
P'_{abc}=L_{aa'}L_{bb'}L_{cc'} P_{a'b'c'},
\label{eq:gaugeon}
\] 
since the transformation can be absorbed by the redefinition $\phi'_a=\phi_{a'} L_{a'a}$. In addition, 
the overall scale transformation of $P$, 
\[
P'_{abc}=e^\psi P_{abc},
\label{eq:gauged}
\] 
with an arbitrary real number $\psi$, does not change the properties of the statistical system,
since this merely changes the overall factor of \eq{eq:znm}.
For example, for $N=2$, these invariances allow one to consider a gauge,
\[
P_{111}=1,\ P_{112}=0,\ P_{122}=x_1,\ P_{222}=x_2,
\label{eq:pgauge}
\]
with real $x_i$. 

\begin{figure}
\begin{center}
\includegraphics[width=7cm]{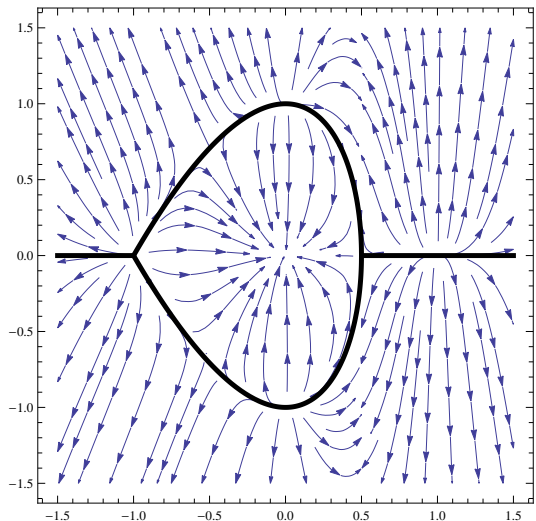}
\caption{The horizontal and vertical axes represent $x_1$ and $x_2$ of 
the gauge \eq{eq:pgauge}, respectively.
The solid lines describe the phase transition lines of the random tensor networks with $N=2$,
which can exactly be obtained in the thermodynamic limit \cite{Sasakura:2014zwa}.
The arrows describe the Hamilton vector flow \eq{eq:hamflow} of CTM with $\mN_a=P_{abb}$.} 
\label{fig:RGflow}
\end{center}
\end{figure}

The free energy per vertex in the thermodynamic limit can be defined by
\[
f(P)= -\lim_{n\rightarrow \infty} \frac{\log Z_{2n}(P)}{2n},
\label{eq:freem}
\]
where we have considered only even numbers of vertices, since an odd number of 
trivalent vertices cannot 
form a closed network. The phase structure of the statistical system can be investigated by
studying the behavior of the free energy \eq{eq:freem}. For the case of $N=2$,
the phase transition lines of the free energy \eq{eq:freem} in the gauge \eq{eq:pgauge} are
shown by solid lines in Fig.\ref{fig:RGflow} \cite{Sasakura:2014zwa,Sasakura:2014yoa}.
The transitions are first order, except at the Curie point, $(x_1,x_2)=(0.5,0)$, where 
the first derivatives of $f(P)$ are continuous, but the second ones are not \cite{Sasakura:2014zwa,Sasakura:2014yoa,revnetwork}.
In fact, for arbitrary $N$, the free energy can exactly be obtained by applying 
the Laplace method to evaluate \eq{eq:freem} \cite{Sasakura:2014zwa,Sasakura:2014yoa}. 
The result is\footnote{If $P$ is symmetric under part of the $O(N)$
transformation \eq{eq:gaugeon}, the perturbations of the integrand of \eq{eq:znm} 
around $\phi=\bar \phi$ contain some zero modes, and the 
application of the Laplace method will require extra treatment to integrate over
the symmetric directions. However, this integration is obviously finite, and will only generate corrections of the 
free energy higher in $1/n$, which do not affect the thermodynamic limit. Thus, the exact free energy in the thermodynamic limit 
is given by \eq{eq:exactf} for the whole region of $P$ including symmetric points.
Note that this argument may change, if one takes a double scaling limit accompanied with $N\rightarrow \infty$, 
which, however, is not considered in this paper.}  
\[
f(P)=\hbox{Min}_{\phi} \, f(P,\phi)= f(P, \bar \phi),
\label{eq:exactf}
\]
where 
\[
f(P, \phi)&=\phi^2-\frac{1}{2} \log \left[ \left(P \phi^3 \right)^2 \right],
\label{eq:exactfbare}
\]
and $\bar \phi$ is defined so as to minimize $f(P,\phi)$ as a function of $\phi$ for given $P$, namely, 
$\bar \phi$ is one of the solutions to the stationary condition,
\[
\left. \frac{\partial f(P, \phi)}{\partial \phi_a} \right|_{\phi=\bar \phi}=
2 \bar \phi_a - \frac{ 3 P_{abc} \bar \phi_b \bar \phi_c}{P \bar \phi^3}=0.
\label{eq:stationary}
\]

The most important implication of the previous paper \cite{Sasakura:2014zwa} 
was that the phase structure of the Ising model on random networks (more exactly, random
tensor networks with $N=2$) in Fig.\ref{fig:RGflow}
can be derived from the Hamilton vector flow of CTM for $N=2$,
if one regards the Hamilton vector flow as a renormalization-group flow, 
as shown in Fig.\ref{fig:RGflow}.
This is surprising, since CTM was proposed aiming for quantum gravity,
and there exist no apparent reasons for CTM to be related to statistical systems on random networks. 
CTM is a totally constrained system with a number of first-class constraints forming 
an algebra which resembles 
the Dirac algebra of the ADM formalism of general relativity 
\cite{Arnowitt:1960es}.
In the classical case, the constraints are given by 
\[
& {\mathcal{H}}_a= \frac{1}{2}P_{abc} P_{bde} M_{cde}, 
\label{eq:hamdef}  \\
&{\mathcal{J}}_{[ab]} = \frac{1}{4} \left(P_{acd} M_{bcd} 
- P_{bcd} M_{acd}  \right), 
\label{eq:jdef}\\
&{\mathcal D}=\frac{1}{6} P_{abc} M_{abc},
\label{eq:ddef}
\]
where ${\mathcal J}$ and ${\mathcal D}$ are the kinematical symmetry generators 
corresponding to the $SO(N)$ \eq{eq:gaugeon} and the scale \eq{eq:gauged} transformations, respectively,
and ${\mathcal{H}}$ and ${\mathcal J}$ may be called Hamiltonian and 
momentum constraints, respectively, in analogy with general relativity \cite{Arnowitt:1960es}.
Here the bracket for the indices of $\mJ$ symbolically represents the antisymmetry, 
$\mJ_{[ab]}=-\mJ_{[ba]}$, and $M$ is the canonical conjugate variable to $P$
defined by
\[
\{ M_{abc}, P_{def} \} 
=\frac{1}{6} \sum_{\sigma}
 \delta_{a\sigma (d)} \delta_{b \sigma (e)} \delta_{c \sigma (f)}, \ \ 
 \{ M_{abc}, M_{def} \}=\{ P_{abc}, P_{def} \}=0, 
 \label{eq:fumpoi}
\] 
where $\{\ ,\ \}$ denotes the Poisson bracket, 
and the summation over $\sigma$ runs over all the permutations of $d,e,f$
to incorporate the symmetric property of the three-index tensors.
The constraints form a first class constraint algebra,
\[
&\{ \mH (\xi^1), \mH (\xi^2) \} = \frac{1}{6}\mJ \left( [\tilde{\xi^1}, \tilde{\xi^2} ] \right), \notag \\
&\{ \mJ (\eta), \mH (\xi) \} = \frac{1}{6} \mH \left(\eta \xi \right), \\
&\{ \mJ (\eta^1), \mJ (\eta^2) \} = \frac{1}{6} \mJ \left( [\eta^1,\eta^2] \right), \\
&\{ \mD, \mH (\xi) \} = \frac{1}{6}\mH (\xi), \\
&\{ \mD, \mJ(\eta) \} =0,
\]
where $\mH(\xi)=\xi_a\mH_a$, $\mJ (\eta)=\eta_{[ab]}\mJ_{[ab]}$ and 
$\tilde{\xi}_{ab}=P_{abc}\xi_c$.

The Hamiltonian of CTM is given by an arbitrary linear combination of the constraints as
\[
H=\mN_a \mH_a + \mN_{[ab]} \mJ_{[ab]}+\mN \mD,
\label{eq:hamofCTM}
\]
where $\mN$'s are the multipliers, which may depend on $P$ in the context of this paper, considering a flow in 
the configuration space of $P$. 
Then, the Hamiltonian vector flow is given 
by\footnote{The direction of the flow is chosen in the manner convenient for later discussions.}  
\[
\frac{d}{ds} P_{abc}=\{ H,P_{abc}\},
\label{eq:hamflow}
\]
where $s$ is a fictitious parameter along the flow.
In the previous paper \cite{Sasakura:2014zwa}, which compares CTM with the random tensor networks 
for $N=2$, 
the multiplier $\mN_a$ is chosen to be
\[
\mN_a=P_{abb},
\]
based on that this is the simplest covariant choice. 
The other multipliers $\mN_{[ab]}$ and $\mN$, related to the symmetry generators,
are chosen so that the Hamilton vector flow \eq{eq:hamflow} keeps the gauge 
condition \eq{eq:pgauge}. Indeed the flow in Fig.\ref{fig:RGflow} has been drawn  
with these choices.  
One can also check that other covariant choices such as $\mN_a=P_{abc} P_{bde}P_{cde}$
do not change the qualitative nature of the flow and therefore 
the coincidence between the phase structure of the random tensor networks and the one implied 
by CTM with $N=2$.

Though the coincidence is remarkable for $N=2$, 
from further study generalizing gauge conditions and values of $N$,  
we have noticed that there exist some problems in insisting the coincidence as 
follows:
\begin{itemize}
\item
First of all, no physical reasons have been given for the coincidence. 
A primary expectation is that there exists a renormalization group procedure for statistical
systems on random networks, 
and the procedure is described by the Hamiltonian of CTM in some manner.
However, it is unclear how one can define a renormalization group procedure for statistical systems 
on random networks, which do not have regular lattice-like structures.
\item
As will explicitly be shown later, in the case of $N=3$, the phase transition lines
deviate from the expectation from the Hamilton vector flow of CTM. 
What is worse is that different choices of $\mN_a$, such as $P_{abb}$ or 
$P_{abc} P_{bde}P_{cde}$, give qualitatively different Hamilton vector flows,
which ruins the predictability of the transition lines from the flow.
\item 
In Fig.\ref{fig:RGflow} for $N=2$, on the phase transition lines, the flow goes along them,
and there exist a few fixed points of the flow on the transition lines.
The fixed point at $(0,1)$ is a co-dimension two fixed point, 
and the associated phase transition is expected to be of second order rather than first order,
if the flow is rigidly interpreted as a renormalization-group flow and we follow the standard 
criterion \cite{renbook}. 
This is in contradiction to the actual order of the phase transition.
This contradiction is more apparent in the diagram in another gauge 
in Section \ref{sec:comparison}. 
\end{itemize} 

The purpose of the present paper is to solve all the above problems, and to show that
CTM actually gives an exact correspondence to random tensor networks. 
It turns out that $\mN_a$ should not be given by any ``reasonable" choices as above, 
but should rather be determined dynamically as $\mN_a \sim \langle \phi_a \rangle$
to be discussed in the following sections. 
Then, we can show that the Hamiltonian of CTM
actually describes a coarse-graining procedure of random tensor networks, 
and that the Hamilton vector flow is in perfect agreement with the phase structure 
irrespective of values of $N$.   

%%%%%%%%%%%%%%%%%%%%%%%%%%%%%%%%%
\section{Renormalization procedure and renormalization-group flow}
\label{sec:renorm}
%%%%%%%%%%%%%%%%%%%%%%%%%%%%%%%%%
\label{sec:ren}
In this subsection, we discuss a renormalization group procedure of 
the random tensor networks, and obtain the corresponding renormalization group flow.

Let us consider an operator $\mO$ which applies on $Z_n(P)$ as  
\[
\mO Z_n(P)=\int d\phi \ \{\phi_a \mH_a ,(P \phi^3)^n\} \, e^{-n \phi^2}.
\label{eq:defofo}
\]
By using \eq{eq:hamdef} and \eq{eq:fumpoi}, and performing partial integrations with respect to $\phi$,
one can derive
\[
\mO Z_n(P)&=\int d\phi \ \{\phi_a \mH_a ,(P \phi^3)^n\} \, e^{-n \phi^2} \CR
&=\frac{1}{2} \int d\phi \ \{\phi_a P_{abc}P_{bde} M_{cde} ,(P \phi^3)^n\} \, e^{-n \phi^2} \CR
&= \frac{n}{2} \int d\phi \ \phi_a P_{abc} P_{bde}\phi_c \phi_d \phi_e (P \phi^3)^{n-1}\, e^{-n \phi^2}\CR
&= \frac{1}{6} \int d\phi \ P_{abc}\phi_a \phi_c\left[ \pder{\phi_b} (P \phi^3)^{n}\right]\, e^{-n \phi^2} \CR
&= -\frac{1}{6} \int d\phi \ (P \phi^3)^{n}\pder{\phi_b} \left[ P_{abc}\phi_a \phi_c e^{-n \phi^2} \right] \CR
&= \frac{1}{3} \int d \phi\  \left[n (P \phi^3)^{n+1} - \phi_a P_{abb} (P \phi^3)^{n} \right]e^{-n \phi^2} \CR
&=\frac{n}{3}\left(\frac{n+1}{n}\right)^\frac{3n+3+N}{2} Z_{n+1}(P)-\frac{1}{3} P_{abb} \langle \phi_a \rangle_n Z_n(P),
\label{eq:operation}
\]
where $\langle \phi_a \rangle_n$ is an expectation value defined by
\[
\langle \phi_a \rangle_n = \frac{ \int d\phi \ \phi_a (P \phi^3)^n e^{-n \phi^2}}{Z_n(P)},
\label{eq:vacuumexp}
\]
and the numerical factor in the first term of \eq{eq:operation} is due to the 
rescaling of $\phi$ for $n \phi^2 \rightarrow (n+1)\phi^2$ in the exponential.

Here \eq{eq:operation} and \eq{eq:vacuumexp} must be used with a caution.
If taken literally, since $Z_{n={\rm odd}}=0$, \eq{eq:operation} and \eq{eq:vacuumexp} do not seem useful by themselves.
The reason for $Z_{n={\rm odd}}=0$ is that the contributions at $\phi=\pm v$ with arbitrary $v$ 
cancel with each other in the integration of \eq{eq:znm}.  
To avoid this cancellation and make \eq{eq:operation} and \eq{eq:vacuumexp} useful, 
let us consider a finite small region $r_{\bar \phi}$ in the space of $\phi$ around 
one of the solutions $\bar \phi$ which minimize \eq{eq:exactfbare}.  
For later convenience, we take the sign of $\bar \phi$ so as to satisfy 
\[
P\bar \phi^3>0.
\label{eq:signofphi}
\]
This can be taken, because, if $P \phi^3=0$, $f(P,\phi)$ in \eq{eq:exactfbare} diverges 
and cannot be the minimum, and $f(P,\phi)=f(P,-\phi)$.
Especially, $r_{\bar \phi}$ should not contain the other minimum $\phi=-\bar \phi$.
Then, let us consider a replacement,
\[
Z_n(P) \rightarrow \int_{r_{\bar \phi}} d\phi \ (P \phi^3)^n \, e^{-n \phi^2}.
\label{eq:repzn}
\]
In the thermodynamic limit $n\rightarrow \infty$, the integral \eq{eq:repzn} is dominated 
by the region with width $\Delta \phi\sim 1/\sqrt{n}$ around 
$\phi=\bar \phi$\footnote{If $P$ is symmetric under part
of the $SO(N)$ transformation \eq{eq:gaugeon}, an extra care will be needed
as discussed in a previous footnote. 
However, this does not change the following argument in the thermodynaic limit.}. 
Therefore, the expression \eq{eq:repzn} approaches $e^{-n f(P)}$ in the thermodynamic limit,
irrespective of even or odd $n$.
Moreover, since the integrand of \eq{eq:repzn} damps exponentially in $n$ on the boundary 
of $r_{\bar \phi}$, the corrections generated by the partial integrations carried out in the derivation of \eq{eq:operation} 
are exponentially small. 
Thus, \eq{eq:operation} is valid up to exponentially small corrections in $n$
after the replacement \eq{eq:repzn}.
Thus, for $n$ large enough, we can safely use \eq{eq:operation} and \eq{eq:vacuumexp}
as if they are meaningful irrespective of even or odd $n$.

Taking into account the discussions above, we can put
$\langle \phi_a \rangle_n \rightarrow \bar \phi_a$ and $Z_n \rightarrow e^{-n f(P)}$ in \eq{eq:operation} for $n \gg 1$.
Therefore, in \eq{eq:operation}, the first term dominates over 
the second term, and one can safely regard $\mO$ as an operator which 
increases the size $n$ of networks. To regard this operation as a flow in the space of $P$
rather than a discrete step of increasing $n$,  
let us introduce the following operator with a continuous parameter $s$,
\[
R(s)=e^{s \mO}.
\]
If we consider $n\gg1$, we can well approximate the operation $\mO$ with 
the first term of \eq{eq:operation} as explained above, and one obtains
\[
R(s) Z_n(P)&=\sum_{m=0}^\infty \frac{s^m}{m!} \mO^m Z_n(P) \CR
&\sim \sum_{m=0}^\infty \frac{s^m (n+m-1)!}{3^m m!(n-1)!} e^{-(n+m)f(P)+\frac{3m}{2}}.
\label{eq:RZn}
\] 
By increasing $s$, the right-hand side is dominated by larger networks, and diverges at $s=s_\infty$,
which is given by
\[
s_\infty=3 \exp\left( f(P)-\frac{3}{2}\right).
\label{eq:sinfty}
\]

On the other hand, in the thermodynamic limit, the left-hand side of \eq{eq:RZn} can be computed in a different manner. 
In the thermodynamic limit, 
$\phi_a$ can be replaced with the mean value $\bar \phi_a$, and the operator $\mO$ can be identified with
a first-order partial differential operator,
\[
\mO\rightarrow \mO_D= \bar \phi_a \mH_a = \frac{1}{2} \bar \phi_a P_{abc}P_{bde} D^P_{cde},
\label{eq:od}
\]
where $D^P_{abc}$ is a partial derivative with respect to $P_{abc}$ with a normalization,
\[
D^P_{abc} P_{def}=\{ M_{abc}, P_{def} \}=\frac{1}{6} \sum_{\sigma}
 \delta_{a\sigma (d)} \delta_{b \sigma (e)} \delta_{c \sigma (f)}.
\]
Here note that $\bar \phi$ is a function of $P$ determined through the minimization of \eq{eq:exactfbare}.
Then, the expression in the left-hand side of \eq{eq:RZn} is obviously a solution to 
a first-order partial differential equation,
\[
\left( \frac{\partial}{\partial s}-\mO_D \right) R(s) Z_n(P) =0.
\label{eq:firsteqzn}
\]
The solution to \eq{eq:firsteqzn} can be obtained by the method of characteristics and is given by
\[
R(s) Z_n(P)=Z_n(P(s)),
\label{eq:ZnPs}
 \]
where $P(s)$ is a solution to a flow equation,
\[
&\frac{d}{ds} P_{abc}(s) =\mO_D P_{abc}(s) \CR
&\hspace{1.7cm}=\frac{1}{6} \left( \bar \phi_d P_{d a e}(s)P_{ ebc}(s)+\bar \phi_d P_{d b e}(s)P_{ eca}(s)
+\bar \phi_d P_{d c e}(s)P_{eab}(s) \right), 
\label{eq:floweq}\\
&P_{abc}(s=0)=P_{abc},
\]
where $\bar \phi$ must be regarded as a function of $P(s)$ by substituting $P$ 
with $P(s)$. 

Here we summarize what we have obtained from the above discussions.
$R(s)Z_n(P)$ can be evaluated in two different ways. 
One is \eq{eq:RZn}, a summation of 
random tensor networks, the dominant size of which increases 
as $s$ increases, while $P$ is unchanged. 
The other is \eq{eq:ZnPs}, where $P(s)$ changes with the flow equation \eq{eq:floweq},
while the size of random networks is unchanged.
This means that the change of the size of networks can be translated into the change of $P$.
Therefore the flow of $P(s)$ in \eq{eq:floweq} 
can be regarded as a renormalization-group flow of the random tensor networks,
where increasing $s$ corresponds to the infrared direction.

\begin{figure}
\begin{center}
\includegraphics{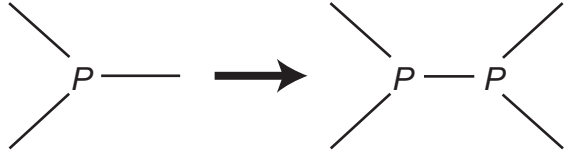}
\caption{The diagrammatic representation of the operation $\{ \phi_a \mH_a, P\phi^3\}$.}
\label{fig:insert}
\end{center}
\end{figure}

The above derivation of the renormalization group flow uses 
the particular form of $\mH$ in \eq{eq:hamdef}. 
Since, in general, there exist various schemes for renormalization procedures for statistical
systems, one would suspect that there would be other possible forms of $\mH$ 
which describe renormalization procedures for random tensor networks. 
However, this is unlikely, and the form \eq{eq:hamdef} would be the unique and the simplest. 
The reason for the uniqueness is that, as outlined in Section~\ref{sec:previous}, the algebraic consistency of 
$\mH$ with the $O(N)$ symmetry, which is actually the symmetry of random tensor networks in the 
form \eq{eq:znm}, requires the unique form \eq{eq:hamdef} 
under some physically reasonable assumptions \cite{Sasakura:2012fb}.
On the other hand, the reason for the simplest can be found by considering the diagrammatic meaning of the operation $\mH$
in \eq{eq:defofo}.  $\mH$ acts on a vertex as
\[
\{\phi_a \mH_a, P\phi^3 \}=\frac{1}{2}\phi_a \phi_b P_{abc} P_{cde} \phi_d \phi_e,
\label{eq:hp3}
\]
and hence can be regarded as an operator which
inserts a vertex on an arbitrary connection in a network (See Fig.\ref{fig:insert}).
This is obviously the most fundamental operation which increases the number of vertices of a network.

Here we comment on our new proposal in comparison with the previous one \cite{Sasakura:2014zwa}. 
Our main claim is that the multiplier should take $\mN_a=\phi_a$ rather than
``reasonable" choices such as $\mN_a=P_{abb},\ P_{abc}P_{bed}P_{cde}$, etc., taken 
in the previous proposal. 
With $\mN_a=\phi_a$, the Hamiltonian vector flow is uniquely determined by the dynamics, while 
``reasonable" choices are ambiguous. 
Even if ambiguous, there are no problems in the $N=2$ case, 
since there are no qualitative changes of the flow among ``reasonable" choices,
and the phase structure can uniquely be determined from the flow.
However, as will be shown in Section~\ref{sec:comparison}, this is not 
true in general for $N>2$.
In fact, $N=2$ is special for the following reasons.   
It is true that $\bar  \phi_a$ can well be approximated by $\sim const. P_{abb}$  near
the absorbing fixed points in Fig.\ref{fig:RGflow},
because all of them can be shown to be gauge-equivalent to $P_{111}=1, \hbox{ others}=0$.
This means that at least an approximate phase structure can be obtained even by putting
$\mN_a=P_{abb}\sim \bar \phi_a$.
In addition, what makes the $N=2$ case very special is that the phase transition lines 
are the fixed points of the $Z_2$ symmetry corresponding to reversing the sign of the magnetic field of 
the Ising model on random networks. Therefore, the phase transition lines are protected by the symmetry, which stabilizes 
the qualitative properties of the flow under any changes of $\mN_a$ respecting the symmetry.
  
Finally, we comment on an equation which can be derived from \eq{eq:operation}
in the thermodynamic limit.
By putting $Z_n(P)\sim e^{-n f(P)}$ to \eq{eq:operation}, one can derive
\[
\bar \phi_a P_{abc}P_{bde} D^P_{cde} f(P)= -\frac{2}{3} e^{-f(P)+\frac{3}{2}}.
\label{eq:eqforfp}
\]
In fact, one can directly prove \eq{eq:eqforfp}.
By using \eq{eq:exactf} and \eq{eq:exactfbare}, the left-hand side of \eq{eq:eqforfp} is given by
\[
\bar \phi_a P_{abc}P_{bde} D^P_{cde} f(P)
&=\bar \phi_a P_{abc}P_{bde} (D^P_{cde} \bar \phi_g) 
\frac{\partial f(P,\bar \phi)}{\partial \bar \phi_g}-\bar \phi_a P_{abc}P_{bde} 
\frac{\bar \phi_c \bar \phi_d \bar \phi_e}{P\bar\phi^3} \CR
&=-\frac{2}{3} P \bar \phi^3, 
\label{eq:pphi3rel}
 \]
where we have used \eq{eq:stationary}.
This coincides with the right-hand side of \eq{eq:eqforfp}, 
because of the choice \eq{eq:signofphi} and 
$\bar \phi^2=\frac{3}{2}$, which can be obtained by contracting \eq{eq:stationary} with $\bar \phi_a$.

%%%%%%%%%%%%%%%%%%%%%%%%
\section{Comparison}
\label{sec:comp}
%%%%%%%%%%%%%%%%%%%%%%%%
\label{sec:comparison}
In this section, we will check the proposal of this paper 
in the cases of $N=2,3$ by comparing with the phase structures of the random tensor networks.

\begin{figure}
\begin{center}
\includegraphics[width=7cm]{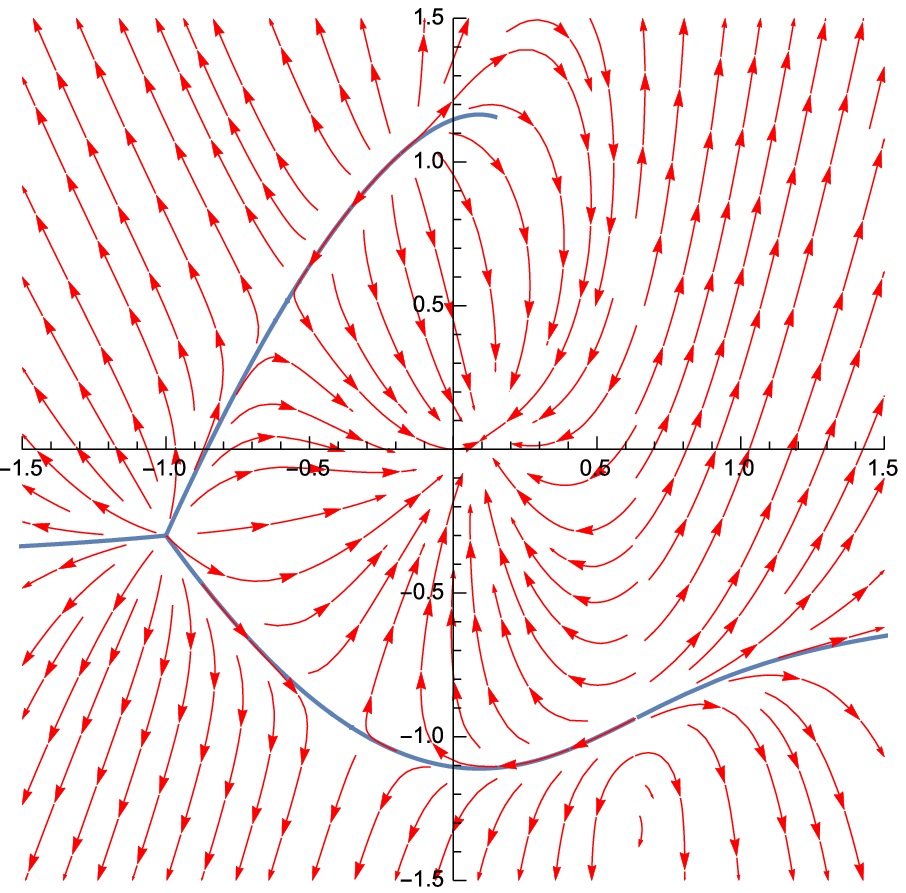}
\hfil
\includegraphics[width=7cm]{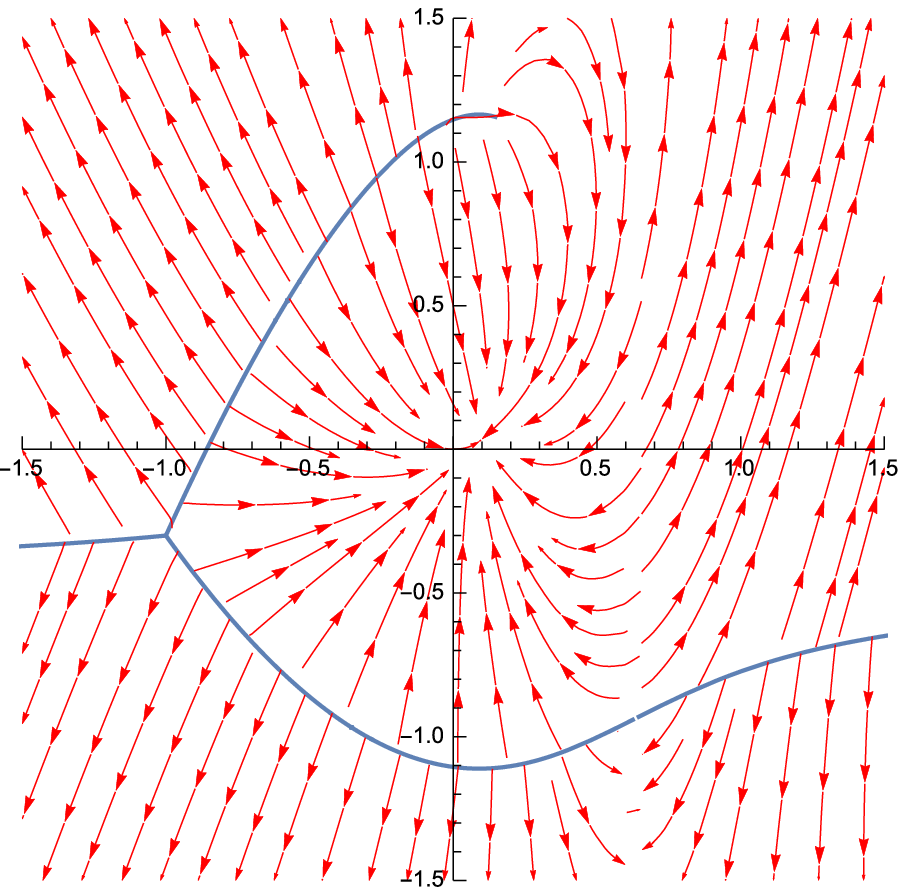}
\caption{The phase transition lines of the random tensor network with $N=2$
in the gauge \eq{eq:diffgauge} are shown as solid lines. 
The horizontal and vertical axes indicate $x_1$ and $x_2$, respectively. 
The phase transitions are first-order except for the endpoint of a line located around $(x_1,x_2)=(0.2,1.2)$. This 
is the Curie point, which is gauge equivalent to $(x_1,x_2)=(0.5,0)$ in the gauge \eq{eq:pgauge}.
The left figure describes the Hamilton vector flow 
based on our former proposal, $\mN_a=P_{abb}$,
while the right figure describes it based on our new proposal, $\mN_a= \bar \phi_a$.
A locus of gauge singularities is located at $x_1=0.635$.
Gauge singularities are not physical, and the free energy has  no singular behaviors there. 
Some details are given in Appendix~\ref{app:n2}.}
\label{fig:phaseN=2}
\end{center} 
\end{figure}

Let us first consider the $N=2$ case with a gauge,
\[
P_{111}=1,\ P_{112}=0.3,\ P_{122}=x_1,\ P_{222}=x_2,
\label{eq:diffgauge}
\]
as a typical example. The difference from \eq{eq:pgauge} is the gauge fixing value of $P_{112}$.
One can obtain the phase structure in the parameter space of $(x_1,x_2)$ 
by studying the free energy \eq{eq:exactf}. 
Alternatively, one can apply the $O(2)$ and scale 
transformations, \eq{eq:gaugeon} and \eq{eq:gauged}, on $P$
so that the phase structure in the gauge \eq{eq:pgauge} 
given in Fig.\ref{fig:RGflow} is transformed to that in the gauge \eq{eq:diffgauge}.
In either way, one can determine the phase structure in the new gauge,
and the result is given in Fig.\ref{fig:phaseN=2}. 
Here we draw the Hamilton vector flows for $\mN_a=P_{abb}$, based on the former proposal, 
and $\mN_a=\bar \phi_a$, based on our new proposal, in the left and right figures, respectively.

The rough features of the two flows based on the different proposals seem consistent with the phase structure: 
the flows depart from the transition lines, and go into the same absorption fixed points.
This was the main argument in our previous paper \cite{Sasakura:2014zwa}.
However, there are some physically important differences in details between the left and the right figures.
In the left figure, on the phase transition lines, the flow is going along them.
Moreover, there exist a few fixed points of the flow on the transition lines 
at $(x_1,x_2)\sim (-0.2,1),(-0.6,0.5),(-0.4,-1)$ in the left figure.  
If the flow is strictly interpreted as a renormalization-group flow, 
the phase transition line on the righthand side of the fixed point near $(-0.2,1)$ is expected 
to be of second order, rather than first order, since the points on the both sides of the transition line
in its vicinity flow to the same fixed point near $(0.1,0)$ without any discontinuities.  
On the other hand, in the right figure,
the flow has discontinuity on the transition lines,  
except for the Curie point at the endpoint of the transition line.  
Thus, the flow based on our former proposal clearly contradicts the actual order of the phase transitions, 
while the one based our new proposal is in agreement with it, i.e. first order except for the Curie point.

An interesting property of the flow is that it does not vanish even on the Curie point,
as can be seen in the right figure of Fig.\ref{fig:phaseN=2} and can also be checked numerically.
This seems curious, because the second derivatives of the free energy 
contain divergences on the point.
In a statistical system on a regular lattice, 
such divergences originate from an infinite correlation length.
Therefore, such a point will typically become a fixed point of a renormalization-group flow. 
On the other hand, 
the correlation length of the Ising model on random networks is known to be finite
even on the Curie point \cite{revnetwork}.
This means that, even starting from the Curie point, a renormalization process will
bring the system to one with a vanishing correlation length. 
This implies that the Curie point cannot be a fixed point of a renormalization-group flow, and 
this is correctly reflected in the fact that our flow does not vanish on the Curie point.
 
\begin{figure}
 \begin{center}
\includegraphics[width=7cm]{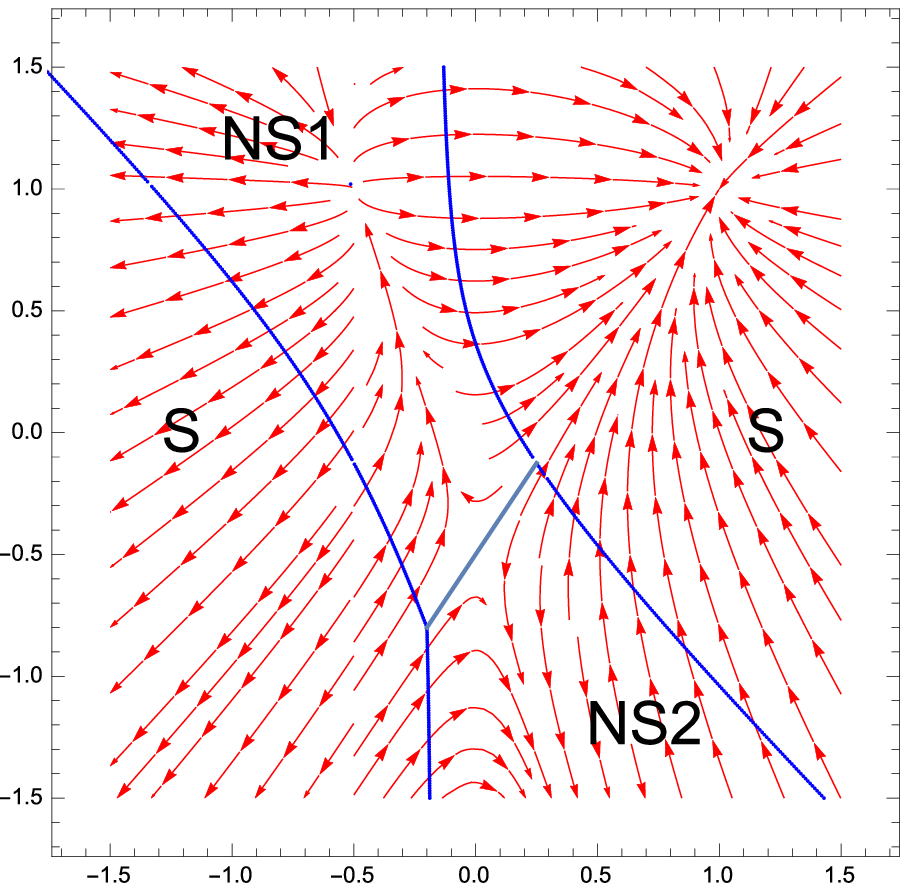}
\hfil
\includegraphics[width=7cm]{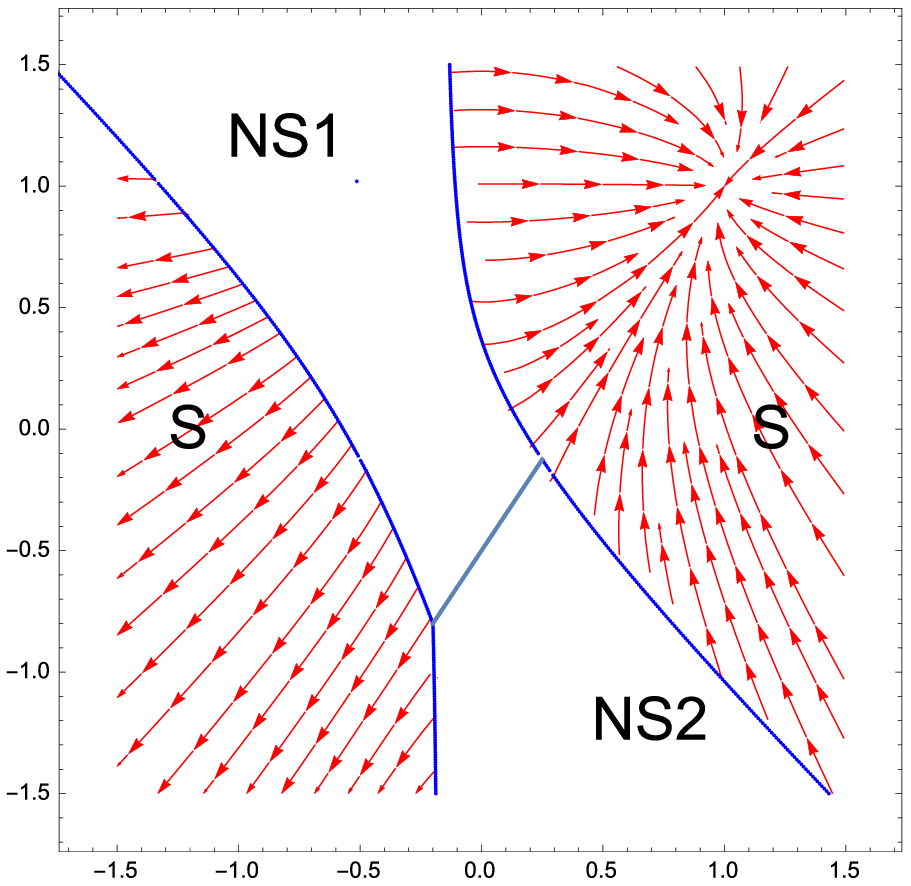}
\caption{The phase transition lines of the random tensor network with $N=3$
in the subspace \eq{eq:subspace} are shown as solid lines. 
The horizontal and vertical axes indicate $x_1$ and $x_2$, respectively. 
There exist four regions separated by the phase transition lines in the indicated parameter region. 
The phases labeled by S have $S_3$ symmetric expectation values 
$\bar \phi_1=\bar \phi_2=\bar \phi_3$. There are two other regions labeled by 
NS1 and NS2, which have non-symmetric expectation values. 
The phase transitions are first-order except for the meeting point of S, NS1, NS2 at 
$(x_1,x_2)= (1/4,-1/8)$. On this point, the free energy is regular in the first derivatives, 
but singular in the second.
On the phase transition line between NS1 and NS2, $P$ is symmetric under 
an $SO(2)$ transfomation \cite{Narain:2014cya}. 
In the left figure, we draw the Hamilton vector flow based on our former proposal $\mN_a=P_{abb}$,
while the right one is based on our new proposal $\mN_a=\bar\phi_a$.}
\label{fig:phaseN=3}
\end{center} \end{figure}

Let us next consider the $N=3$ case. 
There seem to exist too many parameters to treat this case in full generality. 
So let us specifically consider a subspace parametrized by
\[
P_{iii}=1,\ P_{ijj}=x_1,\ P_{123}=x_2,\ \ (i \neq j),
\label{eq:subspace}
\]
which is invariant under the $S_3$ transformation permuting the index labels, 1,2,3. 
Through numerical study of the free energy \eq{eq:exactf} (and some analytic considerations), 
one can obtain the phase structure shown in Fig.\ref{fig:phaseN=3}.
In the indicated parameter region, 
there exist two regions of an $S_3$ symmetric phase labeled by S with 
$\bar \phi_1= \bar \phi_2=\bar \phi_3$. There also exist two distinct non-symmetric phases labeled by 
NS1 and NS2. At any point in the two regions, the minimization of the free energy \eq{eq:exactfbare}
has three distinct solutions of non-symmetric values, $\bar \phi_1\neq \bar \phi_2=\bar \phi_3$,      
$\bar \phi_2\neq \bar \phi_3=\bar \phi_1$, $\bar \phi_3\neq \bar \phi_1=\bar \phi_2$, 
and hence three distinct phases coexist in these regions.
When $S_3$ symmetric subspace \eq{eq:subspace} is extended to more general cases,
each of NS1 and NS2 becomes the common phase boundary of the three phases.

The flow in the left figure of Fig.\ref{fig:phaseN=3} is drawn based on our previous proposal
$\mN_a=P_{abb}$. There, the flow is not in good agreement with the phase structure,
though it seems to capture some rough features.  
We tried other possibilities such as $\mN_a=P_{abc}P_{bde}P_{cde}$, etc., but
the flow depended on the choices, and no good agreement could be found.
On the other hand, in the right figure based on our new proposal $\mN_a=\bar \phi_a$, 
the flow in the symmetric region S is consistent with the phase structure:
the flow departs from the transition lines, and, 
since it does not vanish on the lines, the order is expected to be first order. 
This is in agreement with the property of the free energy except for a point $(x_1,x_2)=(1/4,-1/8)$.
At this point, the free energy is continuous in the first derivatives,
but singular in the second. However, since the flow does not vanish on the point,
the correlation length is expected to be finite. This is similar to the Curie point of the $N=2$ case.

In the non-symmetric phases, NS1 and NS2, the expectation values are not $S_3$ symmetric.
Therefore, the flow has generally directions away from the $S_3$ symmetric subspace,
and cannot be drawn on the figure.
To also check the consistency of the flow in these regions,  
it would be necessary to extend the parameter region.
This would require to take a different systematic strategy for consistency check to avoid too many parameters.
In Section~\ref{sec:theorem}, we will take another way of consistency check by proving a theorem relating 
the renormalization group flow and the phase transitions of the random tensor network.

%%%%%%%%%%%%%%%%%%%%%%%%%%%%%%
\section{Asymptotic behavior of the flow}
\label{sec:fixed}
%%%%%%%%%%%%%%%%%%%%%%%%%%%%%%
In Section~\ref{sec:ren}, we argued that ${\cal O}= \phi_a \mH_a$ provides a 
renormalization procedure for the random tensor network. 
As can be seen from \eq{eq:RZn} and \eq{eq:ZnPs},
$P(s)$ diverges in the limit $s\rightarrow s_\infty$ in \eq{eq:sinfty}. On the other hand, in the numerical analysis
of Section~\ref{sec:comparison}, $P$ is kept normalized as \eq{eq:diffgauge} and \eq{eq:subspace}
by appropriately tuning the multiplier $\mN$ for the scale transformation $\mD$
 (with $\mN_{ab} \mJ_{ab}$ as well). 
As in Fig.\ref{fig:phaseN=2} and \ref{fig:phaseN=3}, one can find fixed points of the flows in 
the limit $\tilde s\rightarrow \infty$, where $\tilde s$ denotes the fictitious parameter parameterizing the 
normalized flows. In this section, we will show that these two limits of $s$ and $\tilde s$ are physically equivalent.

Let us first show the divergence in $s\rightarrow s_\infty$ more directly. 
Since $\bar\phi^2=\frac{3}{2}$ from \eq{eq:stationary}, $\log P\bar \phi^3 =-f(P)+\frac{3}{2}$.
Then, by using \eq{eq:od} and \eq{eq:pphi3rel}, one obtains
\[
\frac{d}{ds} \log P\bar\phi^3=\frac{1}{3} P\bar \phi^3,
\label{eq:difeqpphi3}
\]
where $P$ is meant to be $P(s)$, and hence $P \bar \phi^3$ is regarded as a function of $s$. 
The solution to \eq{eq:difeqpphi3} is 
\[
P\bar\phi^3 = \frac{1}{\frac{1}{\left. P\bar\phi^3 \right|_{s=0}}-\frac{s}{3}},
\label{eq:pphi3s}
\]
which indeed diverges at $s=\frac{3}{\left. P\bar\phi^3 \right|_{s=0}}=s_\infty$.
Since $\bar \phi$ is normalized by $\bar \phi^2=\frac{3}{2}$, the divergence of \eq{eq:pphi3s} 
can be translated to the divergence of $P(s)$ with a behavior, 
\[
P(s)\sim \frac{const.}{s_\infty-s},
\label{eq:behaviorofPs}
\] 
or higher order in the case that some components of $\bar\phi$ vanish in $s\rightarrow s_\infty$.

Now let us compare the two flows, unnormalized and normalized ones. 
For notational simplicity, let us denote the three indices 
of $P_{abc}$ by one index as $P_i$. The flow equations in $s$ and $\tilde s$ can 
respectively be expressed as 
\[
&\frac{d}{ds}P_i(s) =\bar \phi_a(P(s))\, g_{ai}(P(s)), 
\label{eq:difPi}\\
&\frac{d}{d\tilde s} \tilde P_i(\tilde s) =\bar \phi_a(\tilde P(\tilde s))\,  g_{ai}(\tilde P(\tilde s) )
- \mN \tilde P_i(\tilde s), 
\label{eq:diftildePi} \\
&P_i(s=0)=\tilde P_i(\tilde s=0)=P_i,
\label{eq:initialP}
\] 
where $\mN$ generally depends on $\tilde P_i(\tilde s)$, and 
$g_{ai}(P)$ are the quadratic polynomial functions of $P$, which can be read from \eq{eq:floweq}.
The last term of the second equation comes from $\mN\mD$ in \eq{eq:hamofCTM}, and 
is assumed to be tuned to satisfy a gauge condition normalizing $\tilde P(\tilde s)$. 
Here we ignore the $SO(N)$ generators, $\mJ_{ab}$, for simplicity, but
it is not difficult to extend the following proof to include them. 

The physical properties of the random tensor network do not depend on the 
overall scale of $P$. So let us define the relative values of $P(s)$ and $\tilde P(\tilde s)$ as
\[
Q_{i}(s)=\frac{P_i(s)}{P_0(s)},\ \ \ \tilde Q_{i}(\tilde s)
=\frac{\tilde P_i(\tilde s)}{\tilde P_0(\tilde s)},
\] 
where $P_0(s)\ (\tilde P_0(\tilde s))$ is taken 
from one of $P_i(s)\ (\hbox{resp.} \tilde P_i(\tilde s))$, or a linear combination of them. 
From \eq{eq:stationary}, it is obvious that $\bar \phi(P)$ and 
$\bar \phi(\tilde P)$ actually depend only on $Q$ and $\tilde Q$, respectively.
Then, from \eq{eq:difPi}, 
\[
\frac{1}{P_0}\frac{d}{ds} Q_{i}
&=\frac{1}{P_0} \frac{d}{ds} \frac{P_i}{P_0} \CR
&=\frac{\bar \phi_a(Q)\, g_{ai}(P)}{P_0^2}-\frac{\bar \phi_a(Q)\, g_{a0}(P) P_i }{P_0^3} \CR
&=\bar \phi_a(Q) \left( g_{ai}(Q)- g_{a0}(Q) Q_i \right).
\label{eq:flowofQ} 
\]
In the same manner, 
\[
\frac{1}{\tilde P_0}\frac{d}{ds} \tilde Q_{i}=\bar \phi_a(\tilde Q) \left( g_{ai}(\tilde Q)- g_{a0}(\tilde Q) \tilde Q_i \right). 
\label{eq:flowoftQ}
\]
Note that the last term of \eq{eq:diftildePi} does not contribute to the flow equation of $\tilde Q$.
Since the initial condition \eq{eq:initialP} implies $Q_i(s=0)=\tilde Q_i(\tilde s=0)$, and 
the righthand sides of \eq{eq:flowofQ} and \eq{eq:flowoftQ} are identical,
the flow equations, \eq{eq:flowofQ}
and \eq{eq:flowoftQ}, describe an identical flow with a transformation between
the fictitious parameters,
\[
P_0(s) ds =\tilde P_0(\tilde s) d\tilde s.
\label{eq:relstildes}
\]

As discussed above, the typical behavior of $P_0(s)$ is \eq{eq:behaviorofPs}, 
while $\tilde P_0(\tilde s)$ is assumed to remain finite near an absorption fixed point. 
In such a case,  \eq{eq:relstildes} implies
\[
\tilde s \sim -const. \log(s_\infty-s).
\] 
Therefore, the limits of $s\rightarrow s_\infty$ and $\tilde s \rightarrow \infty$ are 
physically equivalent. As can be checked easily, 
this physical implication does not change, even if we consider
the case that $P_0(s)$ diverges with an order higher than \eq{eq:behaviorofPs}. 

To investigate the physical meaning of the fictitious parameter $\tilde s$, 
let us estimate \eq{eq:RZn} near $s\sim s_\infty$.
We obtain
\[
\sum_{m=0}^\infty \frac{s^m (n+m-1)!}{3^m m!(n-1)!} e^{-(n+m)f(P)+\frac{3m}{2}}&
=\frac{e^{-n f(P)}}{(n-1)!} \sum_{m=0}^\infty (m+1)(m+2)\cdots (n+m-1)\left( \frac{s}{s_\infty} \right)^m \CR
&=const. \frac{d^{n-1}}{ds^{n-1}}\sum_{m=0}^\infty \left( \frac{s}{s_\infty} \right)^{m+n-1} \CR
&=const. \frac{d^{n-1}}{ds^{n-1}} \frac{s^{n-1}}{s_\infty-s} \CR
&\sim const. (s_\infty-s)^{-n} 
\]
Then the average size of networks can be estimated as 
\[
\langle n+m \rangle \sim n+s\frac{d}{ds} \log (s_\infty-s)^{-n} \sim \frac{n s_\infty}{s_\infty-s}.
\] 
Therefore
\[
\tilde s \sim const. \log\left( \hbox{Average size} \right).
\] 
This means that $\tilde s$ corresponds to the standard renormalization-group scale parameter often 
denoted by $\log \Lambda$ in field theory. 

%%%%%%%%%%%%%%%%%%%%%%%%%%%%%%%%%%%%%%%%%%%%%%%%%%%%%%%%%%%%%%%%
\section{Discontinuity of the renormalization-group flow and phase transitions}
\label{sec:theorem}
%%%%%%%%%%%%%%%%%%%%%%%%%%%%%%%%%%%%%%%%%%%%%%%%%%%%%%%%%%%%%%%%
In Section~\ref{sec:comparison}, we see that 
the renormalization-group flow has discontinuity on the first-order phase transition lines
in the examples of the random tensor networks.
In this section, we will prove a general theorem on this aspect.

By using the free energy in the thermodynamic limit (\ref{eq:exactf}), 
the stationary condition (\ref{eq:stationary}) and the flow equation (\ref{eq:floweq}), 
we can prove the following theorem.\\
\\
\textbf{Theorem:} The following three statements are equivalent. 
\begin{itemize}
\item[(i)] The first derivatives of $f(P)$ are continuous at $P$. 
\item[(ii)] $\bar{\phi}$ is continuous at $P$.
\item[(iii)] The renormalization-group flow is continuous at $P$. 
\end{itemize}
\textbf{Proof}:  \break
Let us first prove $\hbox{(i)} \Rightarrow \hbox{(ii)}$.   
From (\ref{eq:exactf}), the first derivatives of $f(P)$ are given by
\[
D^P_{abc} f(P) = - \frac{ \bar{\phi}_a \bar{\phi}_b \bar{\phi}_c}{ P \bar{\phi}^3},
\label{eq:df}
\]
where we have neglected the contributions from the $P$-dependence of $\bar \phi$, 
since $\bar \phi$ satisfies the stationary condition (\ref{eq:stationary}).
By contracting a pair of indices in (\ref{eq:df}), one obtains,
\[
D^P_{aab} f(P) = - \frac{\bar{\phi}_a \bar{\phi}_a \bar{\phi}_b}{P \bar{\phi}^3 } =
 - \frac{3\bar{\phi}_b}{2 P \bar{\phi}^3 } ,
\label{eq:paab}
\]
where we have used $\bar \phi^2=\frac{3}{2}$ derived from \eq{eq:stationary}.
Here note that $P\bar \phi^3$ is continuous at any $P$, 
because the free energy $f(P)$ itself in \eq{eq:exactf} is continuous at any 
$P$\footnote{This can be proven by using that $f(P)$ is the minimum of \eq{eq:exactfbare}, 
which is a continuous function of $\phi$ and $P$.},  
and also $\bar \phi^2=\frac{3}{2}$.
Therefore, if (i) holds, \eq{eq:paab} is continuous and hence $\bar\phi$ is continuous; 
$\hbox{(i)} \Rightarrow \hbox{(ii)}$ has been proven.

The reverse, $\hbox{(ii)}\Rightarrow \hbox{(i)}$, is obviously true from \eq{eq:df}.
Therefore, the statements (i) and (ii) are equivalent: $\hbox{(i)} \Leftrightarrow \hbox{(ii)}$.

Next, as for (ii) $\Rightarrow$ (iii), it is obvious that, if $\bar \phi$ is continuous, the renormalization 
group flow (\ref{eq:floweq}) is also continuous.

Finally, let us prove (iii)$\Rightarrow$(ii),
which will complete the proof of the theorem.
To prove this, we will show that there is a contradiction, if we 
assume both (iii) and that $\bar \phi$ has discontinuity on $P$.

Let us suppose that there is discontinuity of $\bar \phi$ at a point $P$. 
Then, from the definition of $\bar \phi$, there exist multiple distinct solutions of $\bar \phi$ 
to \eq{eq:stationary} which give the same minimum of \eq{eq:exactfbare} at $P$.
Let us take any two of them, $\bar \phi^+$ and $\bar \phi^-$. 
As shown above, $P\bar \phi^3$ is continuous at any point, which means 
\[
A\equiv P(\bar \phi^+)^3=P(\bar \phi^-)^3,
\]
where the value is denoted by $A$ for later usage. Here note $A\neq 0$,
since, otherwise, \eq{eq:exactfbare} diverges and cannot be the minimum.
Then, since $\bar \phi^\pm$ both satisfy \eq{eq:stationary},
we obtain
\[
\bar{\phi}^\pm_a &= \frac{3}{2 A} P_{abc} \bar{\phi}^\pm_b \bar{\phi^\pm_c},
\label{eq:id3} \\
\Delta_a &= \frac{3}{2A} 
P_{abc}
\left( 
\Delta_b \Delta_c + 2  \Delta_b \bar{\phi}^-_c
\right),
\label{eq:id2} 
\]
where $\Delta=\bar \phi^+-\bar \phi^-$, and \eq{eq:id2} has been obtained by 
considering the difference of the two equations in \eq{eq:id3}. 
Note $\Delta\neq 0$, if there exist multiplicity of the solutions.

On the other hand, the assumption (iii) and \eq{eq:floweq} imply 
\[
 \left. \frac{d}{ds} P_{abc}\right|_{\bar\phi=\bar\phi^+} 
- \left. \frac{d}{ds} P_{abc}\right|_{\bar\phi=\bar\phi^-} \propto \Delta_d P_{dae}P_{ebc} 
+\Delta_d P_{dbe}P_{eca} 
+\Delta_d P_{dce}P_{eab} =0.
\label{eq:discflow}
\]    
Then, by contracting \eq{eq:discflow} with three $\bar \phi^+$'s or $\bar \phi^-$'s, and using \eq{eq:id3}, 
we obtain
\[
P_{abc} \Delta_a \bar \phi_b^\pm \bar \phi_c^\pm =0.
\label{eq:ppdelp}
\]
Finally, by contracting \eq{eq:id2} with $\Delta$, we obtain
\[
\Delta^2&=\frac{3}{2A} P_{abc}\left(
 \Delta_a  \Delta_b  \Delta_c+
 2 \Delta_a   \Delta_b \bar \phi^-_c\right) \CR
 &=\frac{3}{2A}P_{abc}\Delta_a \left( \bar \phi_b^+ \bar \phi_c^+ - \bar \phi_b^- \bar \phi_c^-\right) \CR
&=0,
\]
where we have used \eq{eq:ppdelp}. This concludes $\Delta=0$, 
which contradicts the initial assumption of the existence of discontinuity of $\bar \phi$.
Consequently, we have proven the equivalence of (i), (ii) and (iii). \rule{5pt}{10pt}\\

By taking contrapositions, a corollary of the theorem is given by\\
\\
\textbf{Corollary 1:} The following three statements are equivalent.
\begin{itemize}
\item[(i)] $P$ is a first-order phase transition point. (Not all of the first derivatives of $f(P)$ 
are continuous.)
\item[(ii)] $\bar{\phi}$ is not continuous at $P$.
\item[(iii)] The renormalization-group flow is not continuous at $P$. 
\end{itemize}
Another corollary of physical interest is  \\
\\
\textbf{Corollary 2:} If $P$ is a phase transition point higher than first-order,  
the renormalization-group flow is continuous at the critical point.\\
\\
The qualitative behavior of the $N=2$ renormalization-group flow shown in the right figure of 
Fig.\ref{fig:phaseN=2} respects the theorem and corollaries as it should be: 
Corollary 1 is realized on the phase transition lines, and Corollary 2 on the Curie point. 

%%%%%%%%%%%%%%%%%%%%%%%%%%%%%%%%
\section{Summary and discussions}
\label{sec:discussion}
%%%%%%%%%%%%%%%%%%%%%%%%%%%%%%%%
In the previous paper \cite{Sasakura:2014zwa}, it has been found that 
the phase structure of the Ising model on random networks (or random tensor 
networks with $N=2$) can be derived from the canonical tensor model (CTM),
if the Hamilton vector flow of the $N=2$ CTM is regarded
as a renormalization-group flow of the Ising model on random networks.
This was a surprise, since CTM had been developed
aiming for a model of quantum gravity in the Hamilton formalism 
\cite{Sasakura:2013gxg,Sasakura:2012fb,Sasakura:2011sq}.  
Considering the serious lack of real experiments on quantum gravity, 
the aspect that CTM may link quantum gravity and 
concrete statistical systems would be encouraging.

The main achievement of the present paper is to have shown that the 
Hamilton vector flow of CTM with arbitrary $N$ gives a renormalization-group
flow of random tensor networks, where the $N=2$ case, in particular, 
corresponds to the Ising model on 
random networks. In the previous paper \cite{Sasakura:2014zwa}, 
we considered the Hamiltonian of CTM, 
$H=\mN_a \mH_a$,  with ``reasonable" choices of $\mN_a$.
Though it was successful in the $N=2$ case, we have shown in this paper that
the previous procedure of taking $H$ does not work for general $N$, 
and have argued that the correct one is given by $H=\phi_a \mH_a$,
where $\phi_a$ are the integration variables for describing random tensor networks. 
Here an advantage of the present procedure is that $H$ is uniquely determined 
by the dynamics of random tensor networks,
but not by the ambiguous ``reasonable" choices of the previous procedure. 
In fact, applied on random tensor networks, $H=\phi_a \mH_a$ is an operator which 
randomly inserts vertices on connecting lines, and therefore it increases sizes of tensor networks.
This provides an intuitive understanding of the role of $H$ as a renormalization process.
We have performed the detailed analysis of the process, and have actually
derived the Hamilton vector flow of CTM as a renormalization-group flow of the 
random tensor network.
In the last section, we have proven a theorem which relates
the phase transitions of the random tensor network and discontinuity of the renormalization-group flow.

The renormalization-group flow which we have obtained has discontinuities 
on the first-order phase transition lines.
However, there is a critical argument on whether a renormalization-group flow has discontinuities 
on a first-order phase transition line \cite{Sokal:1991ui}. 
Since the argument basically assumes a regular lattice-like structure of a system, 
it would be interesting to investigate a similar argument for a system with a random network structure.
The random tensor network would give an interesting playground
to deepen the idea of the renormalization group in a wider situation. 

Finally, let us comment on possible directions of further study,  
based on the achievements of the present paper.
One is the classification of the fixed points of the Hamilton vector flow. This will provide
the classification of the phases and their transitions of the random tensor network.
This would also be interesting from the view point of quantum gravity.
As discussed in \cite{Narain:2014cya}, the physical wave functions of CTM may have 
peaks at the values of $P$ invariant under some symmetries. In general, on such symmetric values of $P$,
$\bar \phi$ may have multiple solutions, and therefore they are phase transition points.
Such interplay between peaks and phase transitions may give interesting insights into quantum gravity.   
Another direction would be to pursue possible relations between the renormalization procedure of 
the random tensor network and that of the standard field theory. 
In fact, the ``grand" partition function \cite{Sasakura:2014zwa}
of the random tensor network
can be arranged to take the form of a partition function of field theory on a lattice 
by considering an index set labelling lattice points and taking $P$ so as to respect locality.
Then, the Hamilton vector flow of CTM may
be regarded as a renormalization-group flow of the standard field theory. It would be 
highly interesting if CTM makes a bridge between quantum gravity and the standard
field theory.

\section*{Acknowledgment}
NS would like to thank the members of National Institute for Theoretical Physics, 
University of the Witwartersrand, for warm hospitality during my visit, where part of 
the present work has been carried out. The visit was financially supported by the Ishizue 
research supporting program of Kyoto University.
YS would like to thank Yukawa Institute for Theoretical Physics 
for comfortable hospitality and financial support, 
where part of this work was done. 
YS is grateful to Tsunehide Kuroki, Hidehiko Shimada and Fumihiko Sugino for useful discussions and encouragement.

\appendix
\section{Explicit expressions of the constraints}
In this appendix, we give the explicit expressions of the constraints, \eq{eq:hamdef}, \eq{eq:jdef}, \eq{eq:ddef},
in the forms used in Section~\ref{sec:comparison} for $N=2,3$.

\subsection{$N=2$}
\label{app:n2}
In a subspace, 
\[
P_{111}=1,\ P_{112}=y,\ P_{122}=x_1,\ P_{222}=x_2,
\label{eq:gengauge}
\]
with fixed $y$, which contains \eq{eq:diffgauge} as a special case, the constraints are given by
\[
&\left( \mH_1, \mH_2 \right) \CR
&\ \ =\frac{1}{6} \Big( 3 (1 + y^2)\pder{P_{111}} +3 (1 + x_1) y \pder{P_{112}}  
+\left( x_1 + 2 x_1^2 + y (x_2 + 2 y)\right)\pder{x_1}+ 3 x_1 (x_2 + y)\pder{x_2}, 
\label{eq:hamN=2exp}
\CR  
&\ \ \ \ \ \ \ \ 
3 (1 + x_1) y\pder{P_{111}} 
+\left(x_1 + 2 x_1^2 + y (x_2 + 2 y)\right) \pder{P_{112}}+ 3 x_1 (x_2 + y) \pder{x_1}
+3 (x_1^2 + x_2^2) \pder{x_2} \Big),\\
&\mJ_{12}\propto -3 y \pder{P_{111}}+(1 - 2 x_1) \pder{P_{112}} +(-x_2 + 2 y)\pder{x_1} + 3 x_1 \pder{x_2}, \\
&\mD \propto \pder{P_{111}}+y \pder{P_{112}}+x_1\pder{x_1} +  x_2 \pder{x_2}.
\]

A Hamilton vector flow is generated by \eq{eq:hamofCTM}, and, for a given $\mN_a$, 
the multipliers associated to the kinematical symmetries, $\mN_{12}$ and $\mN$, can be determined so that the flow stays in the gauge \eq{eq:gengauge}.
Then, since $P_{111}$ and $P_{112}$ are kept constant by such a flow,
determining such a Hamilton vector flow is actually equivalent to
considering $H=\mN_a \mH_a$, where $\pder{P_{111}}$ and $\pder{P_{112}}$
are substituted by solving the linear equations, $\mJ_{12}=\mD=0$. 
Here we do not write the explicit resultant expression of $H$, since it is rather long 
but the procedure itself is elementary. 
An important issue in the procedure is that there exist exceptional points 
characterized by  
\[
3 y^2 +1-2 x_1=0,
\]
where the set of linear equations, $\mJ_{12}=\mD=0$, become singular and cannot be solved 
for $\pder{P_{111}}, \pder{P_{112}}$.
On these points, $\mN_{12}$ and $\mN$ cannot be chosen so that the gauge be kept.
These are the gauge singularities in Fig.\ref{fig:phaseN=2}, 
located at $x_1=0.635$ for $y=0.3$.

\subsection{$N=3$}
The derivation of the Hamilton vector flow in the $S_3$ symmetric subspace \eq{eq:subspace} 
is similar to the $N=2$ case. A difference is 
that, in \eq{eq:hamofCTM},
the multiplier associated to $\mJ_{ab}$ must be set $\mN_{ab}=0$ to keep the $S_3$ invariance. 
For $\mN_a=1$, which is $S_3$ symmetric, one can choose $\mN$ appropriately 
to keep $P_{iii}=1$, and can obtain the Hamilton vector flow as
\[
H_{S_3}=\frac{2 x_1+x_2}{6}
\left( (1 + 3 x_1 - 6 x_1^2 + 2 x_2)  \pder{x_1}+ 6 x_1 (1 - x_2) \pder{x_2} \right).
\]
This is used to draw the Hamilton vector flow in Fig.\ref{fig:phaseN=3}.

\end{document}